\documentclass{article}

\usepackage{arxiv}

\usepackage[utf8]{inputenc} 
\usepackage[T1]{fontenc}    
\usepackage{hyperref}       
\usepackage{url}            
\usepackage{booktabs}       
\usepackage{amsfonts}       
\usepackage{nicefrac}       
\usepackage{microtype}      
\usepackage{lipsum}		
\usepackage{graphicx}
\usepackage{doi}
\usepackage{subfig}
\usepackage{graphicx}
\usepackage{caption}

\usepackage{amsmath}
\usepackage{soul}

\title{Enhancing Security Awareness Through Gamified Approaches}

\author{ \href{https://orcid.org/0000-0003-4079-9243}{\includegraphics[scale=0.06]{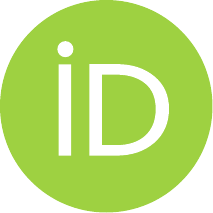}\hspace{1mm}Yussuf ~Ahmed*, Micheal Ezealor, Haitham Mahmoud, MohamedAjmal Azad, Mohamed Ben Farah, Mehdi Yousefi}  \vspace{6pt}  \\
  \vspace{6pt}
	College of Computing, Birmingham City University, Birmingham, UK.\\
	\texttt{Corresponding author email: yussuf.ahmed@bcu.ac.uk. } }

\renewcommand{\headeright}

\hypersetup{
pdftitle={Enhancing Security Awareness Through Gamified Approaches},
pdfsubject={q-bio.NC, q-bio.QM},
pdfauthor={Yussuf.~Ahmmed, Micheal Ezealor, Haitham Mahmoud, MohamedAjmal Azad, Mohamed Ben Farah, Mehdi Yousef},
pdfkeywords={Security Awareness Training; Security Awareness for smart grid, Gamification, Gamification in Cybersecurity 
},
}

\begin{document}
\maketitle

\begin{abstract} With the advent of smart grid (SG) systems, electricity networks have been able to ensure greater efficiency and utility by interconnecting their grids through cloud-based technology. As SGs become increasingly complex, a wide range of security challenges arise, threatening the grid's reliability, safety, efficiency, and stability. The security challenges include the potential exposure of personal data due to hackers intercepting the communications between the SG infrastructure and the smart meters. Security awareness plays a vital role in addressing some of these challenges. However, the traditional training programs are no longer efficient for instilling information security culture in organisations or from an individual user perspective. Gamification is a new concept in the field of information security awareness training (SAT) campaigns that can be introduced to fill in this gap by providing employees with a means of practising and learning about many security flaws and risks that exist within the organisation. Thus, this paper examines the effectiveness of gamification in promoting security awareness among smart meter components for smart grid users/operators. A gaming application is developed as part of the study with the aim of training and evaluating the results through three difficulty levels of questionnaires. Furthermore, the results are evaluated for the three difficulty levels as well as the overall flag captured. It can be demonstrated that the scores of participants in the three levels have improved by 40\%, 35\% and 29\%, respectively. This reflects the awareness of learning within our system.
\end{abstract}

\section{Introduction}

All Industries depend on the goodwill of their consumers and the confidence they inspire in their brands/services. Therefore it is crucial for them to demonstrate security. There may be consequences for many businesses if a major industry suffers a data breach \cite{janakiraman2018effect}. According to a recent report by IBM, the average cost of a data breach is costly and estimated to be around 4.45 million US dollar (USD) \cite{IBM}. The energy sector has been a top target for cybercriminals and accounted for 24\% of all cyber attacks in the UK in 2021 \cite{energyIBM}. Electric utility providers have a whole chain ranging from power plants to smart meter devices installed in people's homes. Cyber attacks can target this chain and disrupt services by targeting the power plants through Denial of Service (DoS) attacks and ransomware, affecting transmission issues by disconnecting the services remotely and targeting the distribution systems in regional hubs. They could also target home networks by exploiting IoT home devices such as smart meters, which could result in the disclosure of personal information and affect availability.

Cybercrime has become a more frequent occurrence, and the evolving nature of attacks has produced a need for regular awareness among users. Many risks exist for organisations, including malware, password theft, network traffic interception, phishing attacks, distributed denial of service (DDoS), zero-day exploits, social engineering, ransomware, waterhole attacks, and Trojan viruses. In this context, the rate at which cybercrime is increasing is concerning. It is estimated that only in the UK, cybercrime costs £27 billion a year \cite{May_2022}.

In general, employees are fundamentally responsible for determining the success or failure of a company. The majority of security breaches occur because employees do not follow protocols due to a lack of motivation, as well as a lack of awareness and knowledge of threats and attacks and insufficient ability to detect and prevent them \cite{abawajy2014user}. 80\% of data breaches, humans are responsible for the breach due to a lack of cyber security skills \cite{Anand_2022}. A Verizon Data Breaches Investigations report published later in 2022 report indicated that 82\% of data breaches resulted from human error \cite{Verizon_Business}.

Keeping consumer information secure is essential for many industries (i.e., energy, retail, etc.) since it ensures that businesses adhere to best practices and protect consumer information \cite{cheung2006understanding}. An unwarranted intrusion into an organisation's sensitive data often results from a security breach, which involves compromising information, programs, networks, and hardware on the computer. Smart grids and meters could introduce data security issues due to the real-time data exchange, which hackers could intercept. Many users rely on apps to interact with their smart meters and track their energy consumption, but these apps could have vulnerabilities and introduce a threat to the chain.

As a means of addressing these challenges, it is imperative to implement Security Awareness and Training (SAT) programs. As part of the SAT program, employees and system users learn about cyber security best practices and gain the necessary skills to protect their data \cite{gjertsen2017gamification}. 

SG cybersecurity studies have extensively examined the power system, and there is a gap in addressing meter data specifically. Hence, this paper focuses on improving security awareness and training among smart meter users in the smart grid. However, this idea can be extended to other industries.  

Specifically, this paper aims to improve security awareness among smart meter users in smart grid systems through gamification. Accordingly, the contribution of the paper can be summarised as follows:

\begin{itemize}
    \item Explore the utilisation of gamification tools and their role towards mitigating cybersecurity challenges and improving user security in smart grid systems with a focus on smart meters.
    \item Implementation of the game for three different levels of difficulties; beginner, intermediate and advanced. 
    \item Evaluation of the three difficulties, beginner, intermediate and advanced, in which each level there are three stages of questions. 

\end{itemize}

The rest 
 of the paper is organised as follows. Section \ref{sec2} presents the background and related work on security awareness and gamification. Section \ref{sec3} aims to propose the methodology by presenting the system setup and app design tool, the questionnaire design with the suggested audience. Section \ref{sec4} discusses the results in terms of the game level and stages, and the overall flag captured. Section \ref{sec5} highlights future research directions, and Section \ref{sec6} concludes the paper.


\section{Background and Related Works\label{sec2}}

The energy sector has been a target for cybercriminals due to the impact it will have on the general population. The national grid is a critical infrastructure, and threat actors, including organised criminals and nation-states, were attributed to such attacks. Some of the most prominent attacks include the attack on the Ukraine power grid \cite{sullivan2017cyber} and the ransomware attack on the colonial pipelines in the 2021 \cite{hobbs2021colonial}.

Energy providers have embraced digitisation, and the connectivity between their systems has increased considerably. For instance, energy providers have been pushing for smart meters with the government's support. For example, 32.4 million smart meters are installed in the UK according to a report released in March 2023 \cite{Smart-meter-stat}. One of the main weakness of SG is the smart meters which are vulnerable to cyber attacks due to the limited memory, consumer app used to interface with the meters, the interconnectivity and the complex infrastructure that support the SG \cite{anzalchi2015survey}.

\subsection{Evolution of Conventional power system to Smart Grid (SG)}

\begin{figure*}[htb]
			\centering
		\includegraphics[width = 14cm]{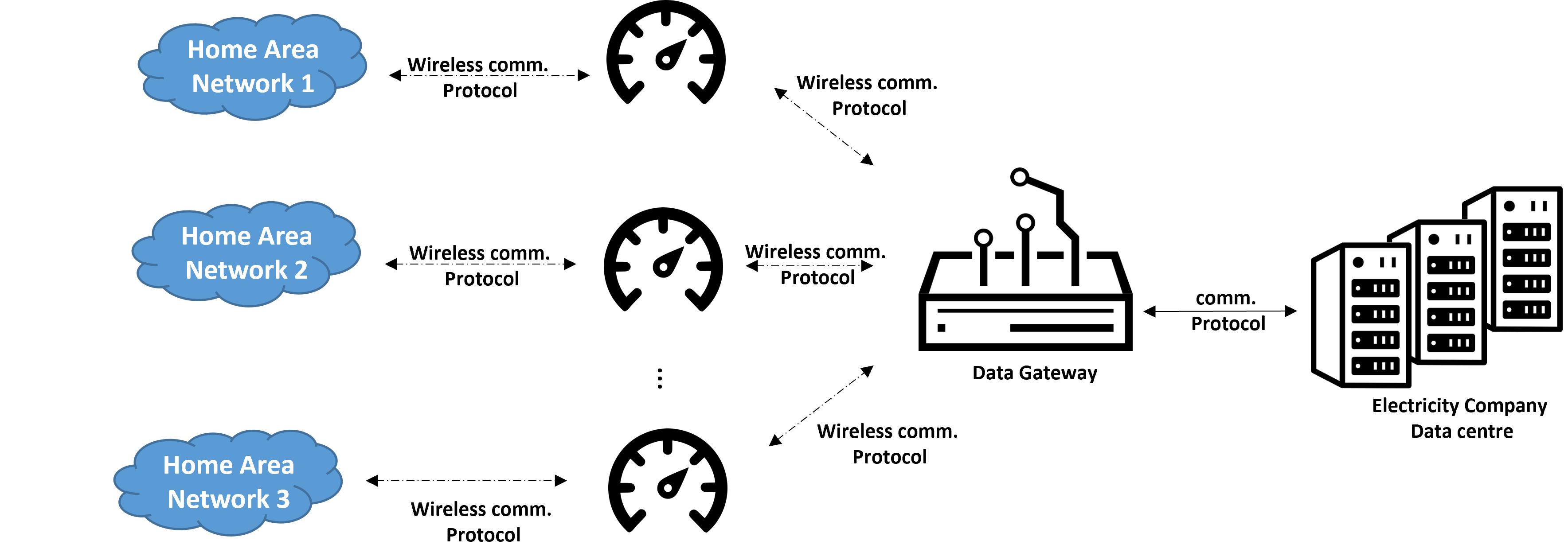}
			\caption{Interfaces of box section of designed system.}
			\label{fig:SC}
		\end{figure*}

A conventional power grid faces numerous challenges, including limited capacity for renewable energy sources, high costs, slow demand response, and carbon emissions \cite{aghaei2013demand}. SG architecture consists of four key components: the home area network, smart meters, data gateway, and the data centre of the electric utility company (see Figure \ref{fig:SC}). Ideally, each household should have a smart meter that transmits its readings to a data gateway using a wireless communication protocol (such as Zigbee), which is then routed to the data centre.

Due to its centralized model, alternative energy sources are difficult to integrate. The system's ineffectiveness can be attributed to the lack of information exchange and bidirectional electrical flow, resulting in manual or semi-automated security and control systems \cite{marin2020security}. 

With the use of SG technology, the conventional power grid brings its limitations to an end by using digital networks and data exchange to overcome those limitations \cite{kabeyi2022use}. It provides improvements in real-time control, operational efficiency, grid resilience, and increased integration of renewable energy, resulting in a reduced carbon footprint due to smart grid implementation. However, there are also significant risks associated with their implementation, including physical and cyber-attacks \cite{zhang2021smart}. The consequences of these risks include the failure of infrastructure, blackouts, the theft of energy, and the compromise of clients' privacy. 

\subsection{Information Security Awareness training}

Organisations have invested heavily in technology-based security measures for many years, but the need to instil an information security culture is ever-pressing. Information security has become the responsibility of everyone in the organisation, and it is no longer left to the information technology departments \cite{ahmed2019cybersecurity}. Employees play a vital role in protecting the organisation, and those with limited or no knowledge of information security awareness are often considered to be the weakest link in the security chain. \cite{dash2022effective}. As risky behaviours rise and networks and Internet applications become more prevalent, information security awareness (ISA) has become an increasingly important topic for both academics and industry experts \cite{parsons2017human, shaw2009impact}. Moreover, Wu et al. (2021) \cite{wu2021assessing} discussed the importance of ISA. During the discussions, they stressed the need for programs designed to increase workers' understanding of information security in order to increase the knowledge and adherence to security practices among workers. Using ISA, health information security regulatory awareness, and punishment severity awareness, Park et al. (2017) \cite{park2017role} developed the framework for health information security awareness (HISA). 

The human factor is a critically important factor in security breaches, particularly in the case of individuals who do not have extensive technical knowledge or are not aware of opportunistic attacks such as social engineering \cite{vroom2004towards, anwar2017gender}. A lack of knowledge about information security can result in non-technical risks and vulnerabilities. These risk factors include user errors, software failures, social engineering, and concerns about data leakage \cite{rezgui2008information}.Information security violations can have serious consequences, such as financial losses and reputational damage. Several human characteristics, including personal norms, self-control, a sense of responsibility and privacy, and attentive awareness, have been associated with security breaches and deviant actions \cite{park2017role, chen2021relationships, alshare2018information}. Taking into account human behaviour plays a vital role in improving security culture and reducing security policy violations within an organisation \cite{rezgui2008information, von2013information}. A defensive security policy must include ISA training and education since individuals are essential in mitigating information security threats \cite{park2017role, stanciu2016students, amankwa2018establishing}.

\subsection{Gamification and Instructional Methods}

Games are used to motivate learning by incorporating mechanisms such as points, badges, levels, and leaderboards into non-game contexts \cite{dicheva2015gamification, chen2020web, barata2017studying}. Gamification aims to engage and motivate students, while serious games deliver educational content, and it can be difficult to distinguish between the two \cite{hamari2014does, marti2016use}. Both concepts refer to gamification \cite{adukaite2017teacher}. 

The use of points, badges, levels, and leaderboards in gamification varies depending on the game mechanics. Points track performance, badges represent achievement, levels correspond to difficulty, and the leader board enables comparison \cite{hamari2014does, buckley2017individualising}.
A variety of game features can be used to provide feedback, set objectives, or recognise the abilities of the user \cite{mekler2017towards, dicheva2015gamification}. 
Badges, goals, and feedback have been shown to significantly enhance behavioural effects in previous studies of the influence of game elements on students' moods and meaningful learning \cite{hamari2017badges}.
The use of badges, leader boards, and performance graphs positively influences employee perception of work meaningfulness and fulfilment of their desire to become competent \cite{sailer2017gamification}. Depending on the duration of the gaming experience, points and levels may be of varying importance \cite{garcia2020perception}. Although badges and leader boards have differing views regarding their effectiveness in motivating employees \cite{hanus2015assessing}.

Gamified education is effective when participants are committed, the educational setting is appropriate, and the time period is appropriate \cite{garcia2020perception}. Gamification fosters positive outcomes, bridges formal and informal learning, and assists with psychological needs \cite{chen2020web, sailer2017gamification}. However, research findings are not always conclusive as to the effects it can have on positive emotions and goal achievement \cite{sanchez2020gamification, acquah2020digital, liao2019interactivity}.

\subsection{Game design Theory}
In this section, two game design theories are discussed: (1) gameplay activities and game mediums, which give examples of various gaming experiences that can be incorporated into training games; and (2) game design approach, which describes the steps taken to guarantee a positive player experience throughout the game development process. 

On one hand, game Play and Medium deal with a variety of gaming activities and media that can be considered when developing a game for training purposes. Various gameplay activities are discussed, including matching, collecting/capturing, allocating resources, strategising, building, solving puzzles, exploring, helping, and role-playing. In addition, board games, role-playing games, exercise, and PC games are mentioned. On the other hand, a game design approach - the player experience design process explains how seven stages of gamification were used to guide the gameplay and player experience. The stages of gamification include outcomes and success metrics, target audience, player goals, engagement model, play space and journey game economy, as well as playtest and iteration.

\subsection{Cybersecurity Awareness via the Use of Gaming Technology}

The rapid adoption of technology is gaining pace. However, users may have a limited understanding of cyber security and security precautions. As a method of education, mobile games can enhance learning activities, provide realistic simulations, and trigger users' creativity. The potential for gaming in learning has been recognised; however, more research is necessary in this area. Mobile gaming education requires users to carry out various learning tasks by using a game or series of games built for a particular learning activity \cite{coenraad2020experiencing, filipczuk2019using}. 

The digital games that may be played on mobile devices have the potential to effectively provide realism of simulations and problem-solving activities, which might boost the learning activity and inspire users. Gamification may provide an essential component of the learning process in the future \cite{zainuddin2020impact}. Several studies have concluded that gaming technology has significant potential for use in a variety of applications. It was shown that the learning process might be considerably enhanced by mobile gaming education. A study by \cite{alotaibi2016review, quayyum2020cyber} focused on the use of gaming for educational purposes. Another study focuses on the gaming process has the potential to raise children's reading skills \cite{camilleri2019mobile}.

Puzzle games are among the most popular and effective educational tools in recent years. They offer an interactive and entertaining approach to instruction, facilitating skill-building and cognitive development. Game techniques can be applied to various training topics, allowing players to make connections between virtual and real-world environments. Games-based learning strategies encourage players to achieve objectives and observe the consequences of their actions \cite{khan2022game}.

The most apparent benefit is that the games-based learning technique offers an interactive approach to instructing or educating users about a particular software application. In an entertaining and participatory manner, it facilitates skill-building and the development of mental processes in the participants. Because of the versatility and flexibility of game techniques, creating a game that applies to practically any potential training topic and targeted at their technical skills \cite{hart2020riskio} is feasible. An effectively developed game will allow the player to enter a virtual environment analogous to the actual environment. This will allow the player to connect what they learn in the virtual environment and what they experience in the real world. The strategy based on games will encourage the player to advance towards the objective by doing the necessary activities, and the player will also be able to observe the effects of their actions without being forced to deal with them in real life. 

However, educating professionals about internal risks remains an untapped area of research. Various applications must be analysed in order to determine their appropriateness and usability. In the past few years, there have been only a few studies examining applications such as Control-Alt-Hack and Anti Phishing Phil and their variations (\cite{awojana2019overview, brady2022gamification, wen2019hack}. There has been positive feedback regarding these studies, indicating that games can effectively raise awareness about cyber security. 

Studies such as "Anti Phishing Phil" and "Security Games by Next Generation Security (NGSEC) have demonstrated that employing games for the goal of learning may result in considerable gains; nevertheless, the sample sizes employed in these studies were rather small. While users of other games have provided favourable feedback, the impact and impacts of such games on the players' learning outcomes have not been evaluated.

\subsection{CyberSecurity, Gamification, and Smart Grid}

A different aspect of smart grid cybersecurity and training is examined, each with a unique focus, a unique benefit, and a different consideration. Two major directions have been taken in the literature on smart grid cybersecurity and training. The first direction involves analysing the information framework to determine whether awareness training is needed \cite{curtis2015evaluating, rob2014addressing}, while the second direction involves building testbeds to assess risks and recommend training \cite{strasser2014co, hahn2013cyber, stites2013smart, jauhar2015model, holm2013cyber}. 
On the one hand, an assessment of cybersecurity capabilities in the energy industry is facilitated by the Cybersecurity Capability Maturity Model (C2M2)\cite{curtis2015evaluating}. Training and awareness are emphasised as recommendations for improvement. By using maturity levels, the model evaluates security and training operations which point out to provide more interactive content to keep user-focused. Moreover, Rob et al. (2014) \cite{rob2014addressing} emphasise the need for cybersecurity solutions in the energy industry, emphasising the importance of a robust internal policy framework, well-trained personnel, and effective awareness programs. Their findings relate to issues such as software selection, delivery methods, ongoing evaluations, and utilising advanced technologies.

On the other hand, a simulation platform is proposed for training in the development of smart grid applications \cite{strasser2014co}. By breaking down the grid system into components and integrating domain-specific tools for simulation and control, the team integrates domain-specific tools for simulation and control. Through the use of simulation platforms, operators are better able to understand each other and have the potential to be utilised in computer security training by introducing new behaviours and simulating attacks. Moreover, the PowerCyber testbed is used for a variety of purposes, including education, training, vulnerability assessment, and evaluating the impact of security threats \cite{hahn2013cyber}. This testbed allows for the exploration of specific attacks and the use of both cyber and physical methods to mitigate their impact.
Furthermore, ISAAC, a smart grid cybersecurity testbed developed by the CPS, is evaluated as a teaching tool for educational purposes \cite{stites2013smart}. Using ThunderCloud as a virtual testbed, the company proposes remote access and a variety of training exercises that simulate real-world vulnerabilities and attacks. Despite increased awareness and preparedness reported by students using the testbed, further evaluation is necessary to confirm these findings. Furthermore, a model-based method, CyberSAGE, is developed for analysing security risks in NESCOR failure scenarios \cite{jauhar2015model}. Using their approach, scenarios requiring technological solutions or personnel training can be identified. Specifically, they found that training on the secure networking requirements reduced failure probability for certain attacker settings but was less effective for physical access scenarios. Finally, a phishing exercise was conducted to assess smart grid security awareness \cite{holm2013cyber}. Several of their study's findings indicated a need for more education regarding the reporting of suspicious emails and security breaches. Based on the findings, assessment activities may be designed following the training session.

However, Smart grid cybersecurity studies have extensively examined the power system, and there is a gap in addressing meter data specifically (See Table \ref{tab1}). Hence, we focused on this paper to improve security awareness and training among smart meter users in the smart grid. 
Smart meters can be used by electricity companies to increase cybersecurity practices, empowering employees to identify and respond to potential threats. Smart meters' integrity and reliability are maintained by adopting a proactive approach that mitigates security risks.

\begin{table*}[ht] 
\centering 
\caption{Related Studies.} \label{tab1}
\begin{tabular}{|l|l|l|}
 \hline
 Ref. & Testbed & Purpose \\
 \hline

\cite{strasser2014co} & NA & Integrating Cyber and Physical components. \\ 
\hline
\cite{hahn2013cyber} & PowerCyber & Vulnerability assessment \& security threats.\\
\hline
 \cite{stites2013smart} & ISAAC & SG tool for educational. \\
 \hline 
\cite{jauhar2015model} & CyberSage & NESCOR failure scenarios. \\
\hline
\cite{holm2013cyber} & NA & Phishing Exercise for SG. \\
\hline
This paper & Root the box & Towards secure smart meter for SG.\\

\hline
\end{tabular}
\end{table*}

\section{Proposed Methodology\label{sec3}}


Our technique is similar to a unique iteration case of the Design Science Research Process (DSRP) \cite{peffers2020design}. Employing a gamified learning application prototype as the medium for engagement can create a more attractive learning atmosphere for security training, providing additional motivation for practising good security behaviour. Our exploratory method is validated by Lebek et al. (2014) \cite{lebek2014information}, who conducted a review of literature on security consciousness and behaviour. Since the quantitative approach dominates the field, they argue that qualitative study could bring value, emphasising the uncommon implementation of such experimental research. The entire system was conceived to be a gaming platform having several elements that may trigger attraction, create fun and cultivate a self-learning experience for users. The platform would be fun-driven rather than appearing to be a learning centre. However, the process will be designed to help retain information. 

\subsection{System Setup and App Design tool}

\begin{figure}[htb]
			\centering
		\includegraphics[width = 8cm]{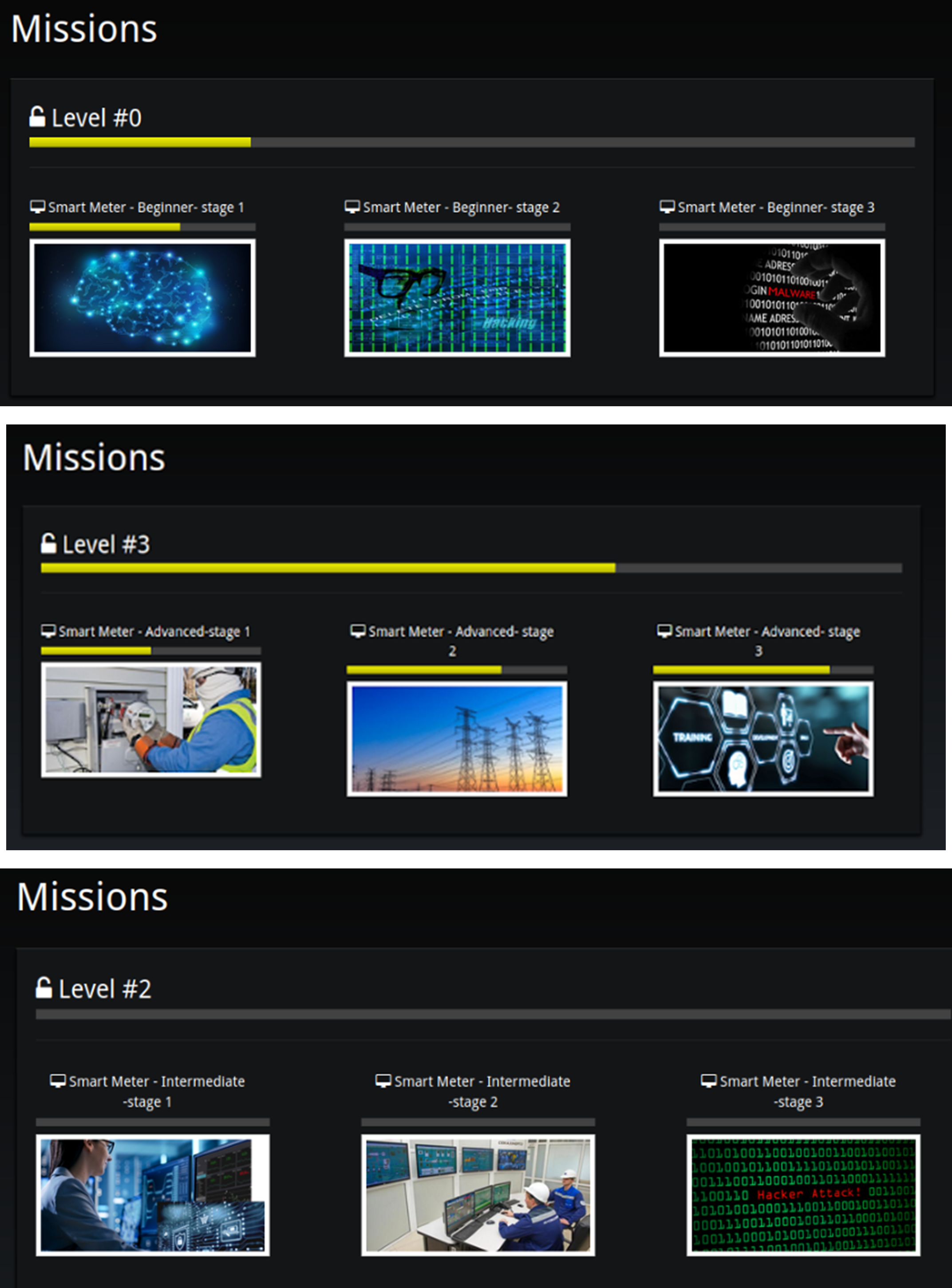}

			\caption{Interfaces of box section of designed system.}
			\label{fig:1}
		\end{figure}

\begin{figure}[htb]
			\centering
		\includegraphics[width = 8cm]{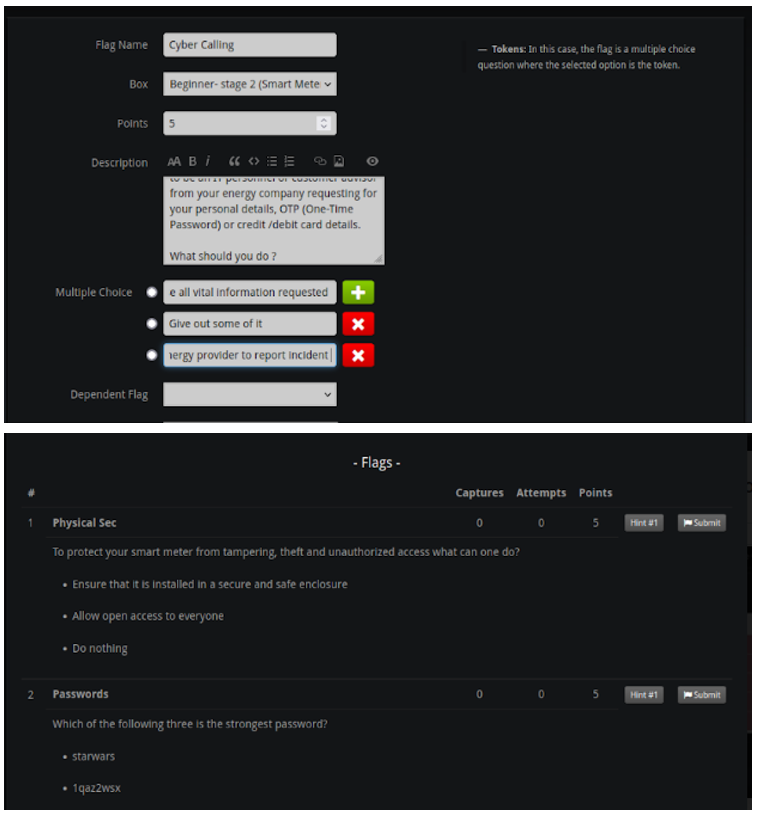}
			\caption{Interfaces of Question flag section of designed system.}
			\label{fig:2}
		\end{figure}

The application prototype was built on software designed for building a gaming application called Root the Box \cite{RoottheBox}. The software offers hackers and developers to hone their skills in simulated computer wargames as well as users' gaming content creation. Root the Box has a real-time capture-the-flag (CTF) scoring engine. The program is highly flexible and can be tailored to fit the needs of any Capture the Flag game. By fusing a game-like setting with genuine tasks that teach skills that are relevant in the real world, including vulnerability assessments, incident handling, computer forensics, and malware detection, the platform may attract players of all skill levels \cite{son2019real}. Root the Box has some tools which make games interactive, some of these tools were engaged in the implementation of the prototype, and they include:

\begin{itemize}
    \item Corporations (Groups): An optional corporation is a collection of containers. A company, for instance, might have flags from both a web server and a database server. In light of this, it is suggested that managers set up their difficulties in a strategic manner (e.g. re-using passwords between boxes within the same corporation). For corporate challenges, it's acceptable to team up multiple boxes or even a complete network of machines.

    \item Boxes (Sections): An information-gathering team's "box" is a grouping of flags that stands for a certain category of data (see flags for details). Each corporation's Box stands in for a host computer and can be attacked by competing groups. Alternately, each cell could stand for an inquiry regarding a specific piece of digital evidence, such as an image memory analysis. Examples of the box section interfaces can be seen in Fig. \ref{fig:1}. 

    \item Flags (Questions): A flag can be a piece of information or a question that the players need to answer to advance. When a player is able to capture a flag, they earn money or points. Some flags are used as hints, while others carry penalties. Examples of the flag interfaces can be seen in Fig. \ref{fig:2}.

    \item Game Levels: The default level for players to start is basic. Capturing a set amount of flags on the current level, spending real money, reaching a certain score, or having the administrator grant access to further levels are all viable options. Unavailable levels cannot be flagged or seen by players. For the prototype, the levels are basic, intermediate and advanced. Each will be accessible depending on the player's level of expertise in the field. 

\end{itemize}

  Our goal in designing a game-like learning application is to increase positive engagement with security best practices. A working, interactive prototype was created to test this theory. All of the security education and practice activities in the prototype software are presented in a gamified format, complete with points, levels, badges, and leaderboards. The players will have control over their login details as they also determine when they access the platform through a web browser. This followed by broad parameters of the applications are described as follows: 

\begin{itemize}
    \item The quizzes would be succinct and brief, and there would be a wide variety of them covering various aspects of smart meter security, with the latter classified into beginner, intermediate, and advanced levels. Answering each question should take no more than two minutes.

    \item In this game, the participants can do the exercises in whatever order they like. This allows for a more emergent model of player involvement, in which they are not required to adhere to a predetermined course of action, and greater freedom for the player. However, limitations must be imposed to guarantee the player receives the necessary instruction.

    \item The prototype will be used twice; once to assess the present skill level of players, then to teach them as they play the game, and once to assess the effect of gamification on their learning.

\end{itemize}

\subsection{Questionnaire Design and Audience}

\begin{figure}[htb]
			\centering
		\includegraphics[width = 8cm]{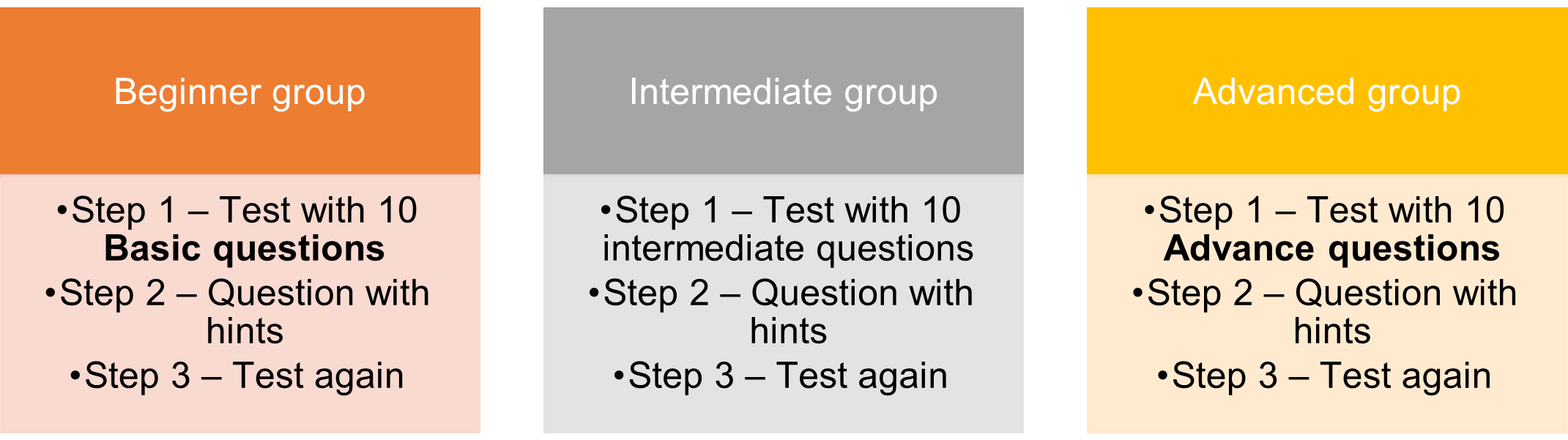}
			\caption{Difficulty levels of the Questionnaire design of the suggested system.}
			\label{fig:3}
		\end{figure}

The quiz contains a series of scenario-based questions ranging from simple ones for beginners and more advanced ones for experts. The questions are designed as multiple choice questions (MCQ), and each addresses the various aspect of smart meter vulnerability based on different individual interactions from consumers to the power company. Therefore, the ideal audience for the survey would be of three categories: individuals who use smart meters and have a basic idea of the system, individuals who are a bit more exposed to the industry either by training, academics or profession. Then the last set of audience will be individuals who have a closer interaction with the system and have acquired an advanced knowledge of how it works, and are conscious of its security vulnerabilities. Three difficulty levels are designed, each with 10 questions of the same difficulty level as seen in Fig. \ref{fig:3}.

The implemented detailed design and development of the game interface and walkthrough for our study are covered in this section of the dissertation. Content building and user interaction play a huge role as they take on the different security-related gaming and gain awareness. The author gives an overview with clear directions and a walkthrough as a user toward an efficient gaming experience.

\begin{figure}[]
\centering
\subfloat[User registration form]{\label{fig:4a}{\includegraphics[width=0.45\textwidth]{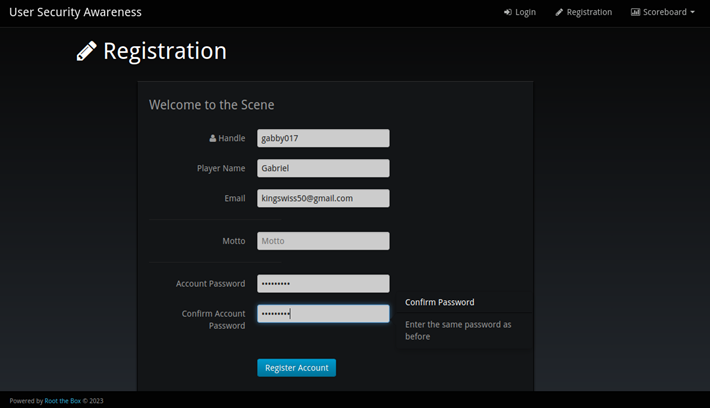}}}\hfill
\subfloat[User completion progress bar]{\label{fig:4d}{\includegraphics[width=0.45\textwidth]{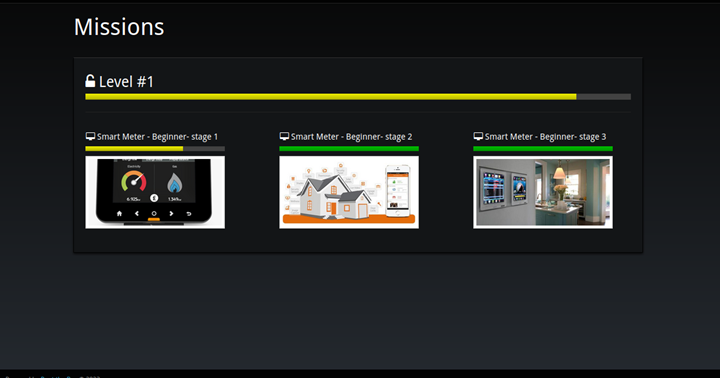}}} \hfill
\subfloat[Gaming question interface]{\label{fig:4e}{\includegraphics[width=0.45\textwidth]{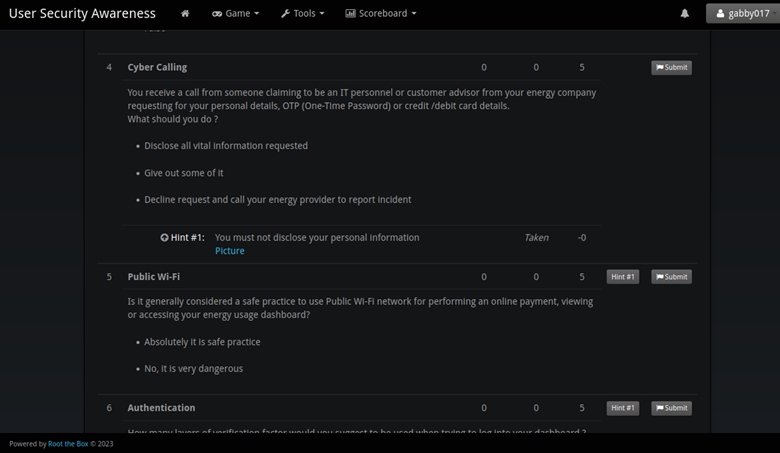}}} \hfill
\subfloat[User progress view]{\label{fig:4g}{\includegraphics[width=0.45\textwidth]{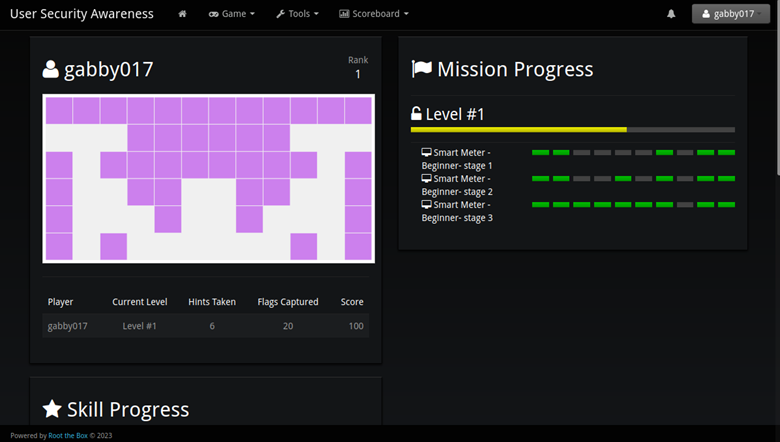}}} \hfill

\caption{Examples for the gaming interface of the designed system.}
\label{fig:4}
\end{figure}

Fig. \ref{fig:4a} shows the registration page of the game, which every user must go through. After a user has successfully registered and logged in, a landing page that welcomes them and provides them with a personalised storytelling and journey introduction is encountered.
As mentioned earlier, every level has three stages based on users’ knowledge level. Various mission stages are implemented by the author and a progress bar to show users work done. As the user plays the game, the progress bar moves further, as seen in Fig. \ref{fig:4d}. The questions implemented to test beginner users alongside other gaming key elements are discussed within the game as seen in Fig. \ref{fig:4e}.
During the game, users are educated to improve their knowledge of security threats and attacks. When the game ends, users can see their performances, rewards, scoreboards and progress achieved. These gaming components are keys as they drive user engagement and enable retention. Fig. \ref{fig:4g} shows user progress, score, capture counts and complete view.

\subsection{Evaluation}
To adequately evaluate the effectiveness of the gamified security awareness process for SG users, the web-based game prototype with individuals has been experimented with through an intermediate to advanced understanding of smart grid tools and functionality, as discussed above. The game application is developed using Root the Box for web browsers first to determine their level of understanding of security issues and potential risks that pertain to the use and operation of the smart grid system. Secondly, to train the players in the areas of potential risks and major security issues, then thirdly, to determine the effectiveness of the training in educating the users. 

Each stage of the evaluation will involve answering a series of questions focusing on the main technical threats and cyber-attacks that are common to the systems. At each stage of the evaluation process, the players will be required to use the gaming application to answer 10 questions each while the necessary insight will be measured. Each question will be multiple choice, including a simplistic multiple-choice quiz and a 5-point Likert scale. The gamification features to be integrated may include specific themes, on-screen progress/feedback, time pressure, consequences and competition. Following the evaluation process, insights will be gathered and reported as evidence to address the aim of the study. 

\section{Results and Discussion\label{sec4}}

The prototype developed in this paper aimed at testing the knowledge of the players on several levels of security awareness of individuals. The system contains questions which range from beginner, intermediate to advance levels and within each level, there are three stages of questions. Respondents were required to play the game and their results are presented in the section

\subsection{Game Level and Stages}

By using this method, we are able to identify the level of difficulty of the question, differentiate the target audience, and reflect on their level of knowledge and interaction with the smart meter system. At the beginner level, individuals with a basic understanding of the subject, primarily homeowners, are enrolled. Intermediate-level courses are intended for those with a mid-level understanding of technicalities and vulnerabilities. This advanced level is designed for those who have a close working relationship with the smart meter system and have an in-depth understanding of the technical aspects and key vulnerabilities.

Moreover, there is a progressive level of difficulty within each level, separated into stages. As a result of this approach, players are able to adapt gradually to the game mechanics and obstacles rather than being overwhelmed by them all at once. This is a common method in games that have a learning curve, as it introduces new players to the game and allows them to gain some familiarity with it. Players benefit from this strategy by maintaining their interest and feeling accomplished as their skills improve. Every level consists of three phases: assessment, training, and evaluation. During each stage, there are ten questions covering a variety of topics. Players are provided with hints during the training phase to assist them in finding the answers they seek. The findings of each level of respondents are presented here and in the following discussion.

\subsubsection{Beginners-level Evaluation}

These questions are targeted towards people with little or no cyber security knowledge utilising information. In our case study, we targeted smart meter owners that are typical users that own and use smart meters. On the chain of stakeholders of the smart metering system, they are referred to as consumers.

\begin{figure*}[]
\centering

\subfloat[User's score of the three stages]{\label{fig:5a}{\includegraphics[width=.6\textwidth]{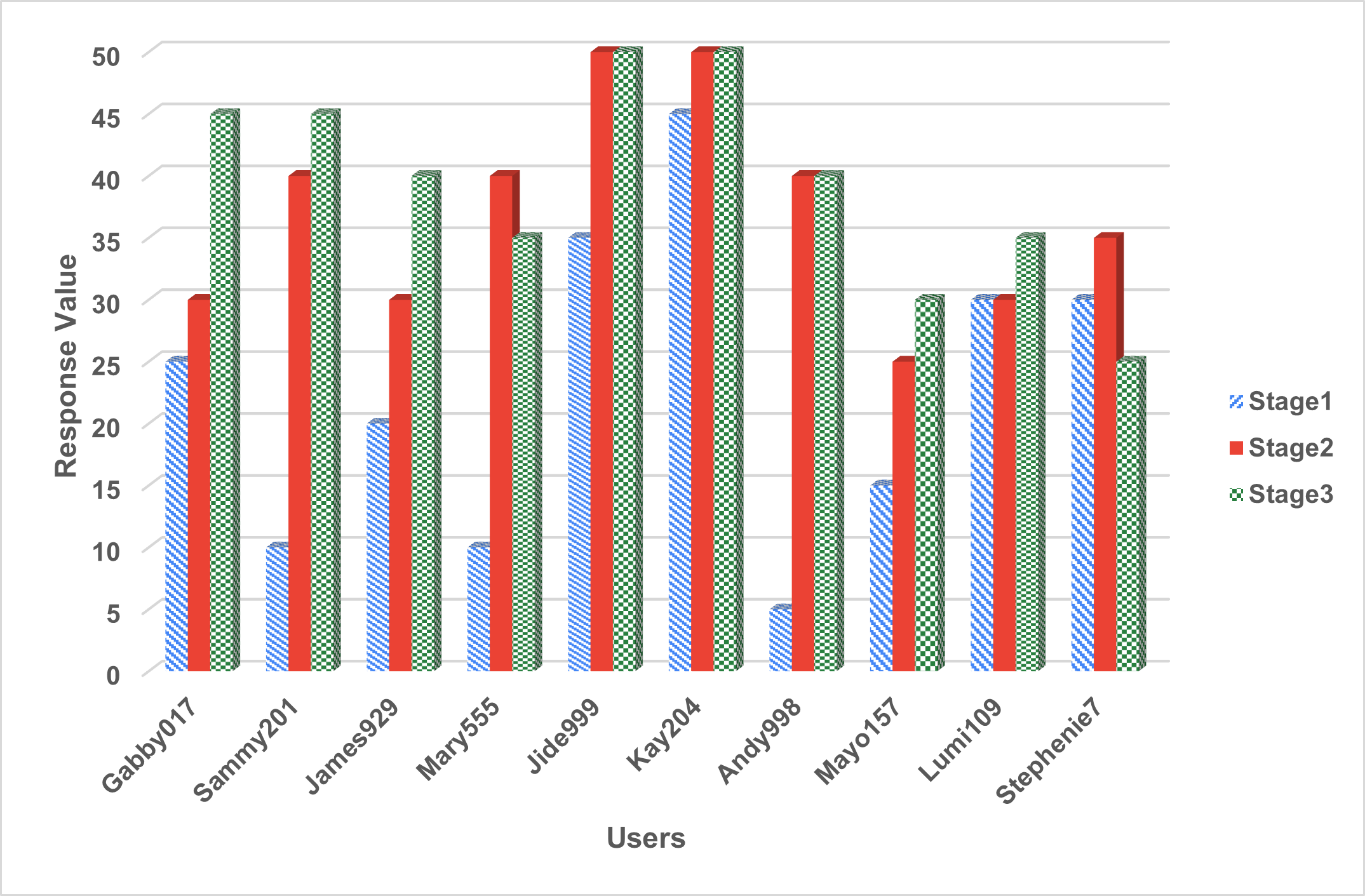}}}\hfill

\subfloat[No. of Hints taken]{\label{fig:5b}{\includegraphics[width=.6\textwidth]{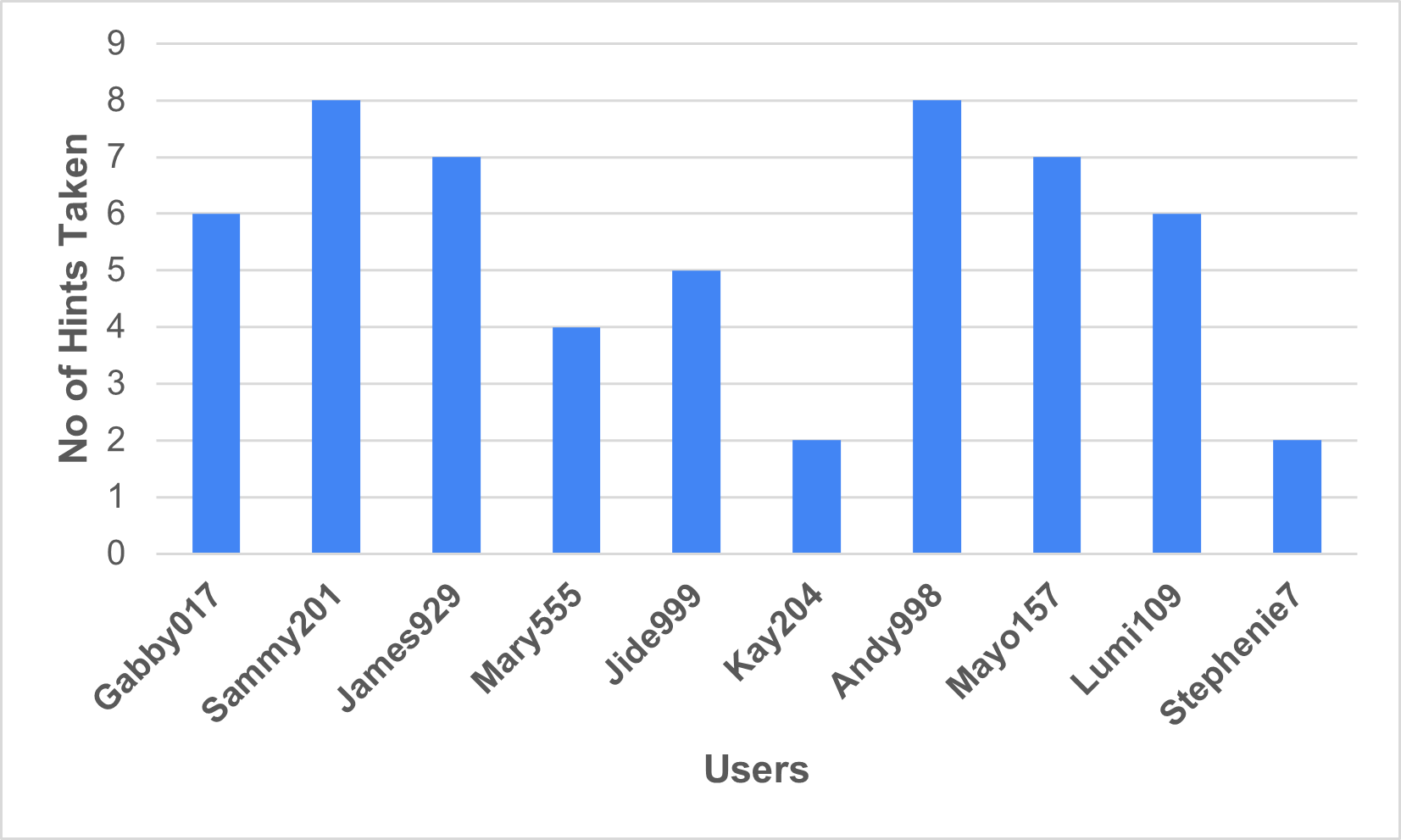}}}\hfill


\caption{Beginner level Evaluation of the designed system.}
\label{fig:5}
\end{figure*}

Fig. \ref{fig:5a} shows the overall scores of all ten players at the beginner level for the three stages. Based on the maximum score of 50, it is clear that players have varying levels of knowledge about security issues with smart meters at the basic level. Consequently, it is justified to conclude that the study is necessary and applicable. Kay204 obtained the highest score (45 marks), suggesting that he is already familiar with the issues and thus, gamification may not be easily noticed. Andy998 earned the lowest score (5 marks), followed by Sammy201 and Mary555 with ten points each. The present scenario provides an excellent opportunity to examine the impact of gamification.

The performance of all participants improved in stage two as they had clues to help them understand the question or select the correct response. Two players (Kay204 and Jide999) were able to score 50 using the hints provided. Kay204 scored 45 points in stage one, and improved to 50 points here, and scored 35 points in stage one and improved to 50 points in stage two. The lowest scorers in stage one, Andy998, Sammy201, and Mary555, all improved significantly to 40 marks in stage two.

As seen in Fig \ref{fig:5b}, most of the players engaged with the training technique by taking hints. The rationale behind the hinting in teaching the participants is that taking more hints suggests that they were responsive to learning to discover solutions, while taking fewer hints implies that they did not use the training, which negates the goal of the study. As all players took hints, we can observe how the hint affected their performance in stage three, taking into account their initial score in stage one.
Gamification is proving to be highly effective in training players, as evidenced by the performance in the third stage. There are several notable players: Andy998 scored five points (only one question correctly) in stage one, took eight hints in stage two, thereby improving his score to 40 points in stage two, and then scored 40 again in stage three. During stage one, Sammy201 scored 10 points; in stage two, he took 8 hints which improved his score to 40 points; and in stage three, he scored 45 points. As a result of Jide999's performance, he scored 35 marks in Stage One, 50 points in Stage Two with five hints, and 50 points in Stage Three. At the beginner level, these performances demonstrate the effectiveness of the techniques.  

\subsubsection{Intermediate-level Evaluation}
\begin{figure*}[]
\centering

\subfloat[User's score of the three stages]{\label{fig:6a}{\includegraphics[width=.6\textwidth]{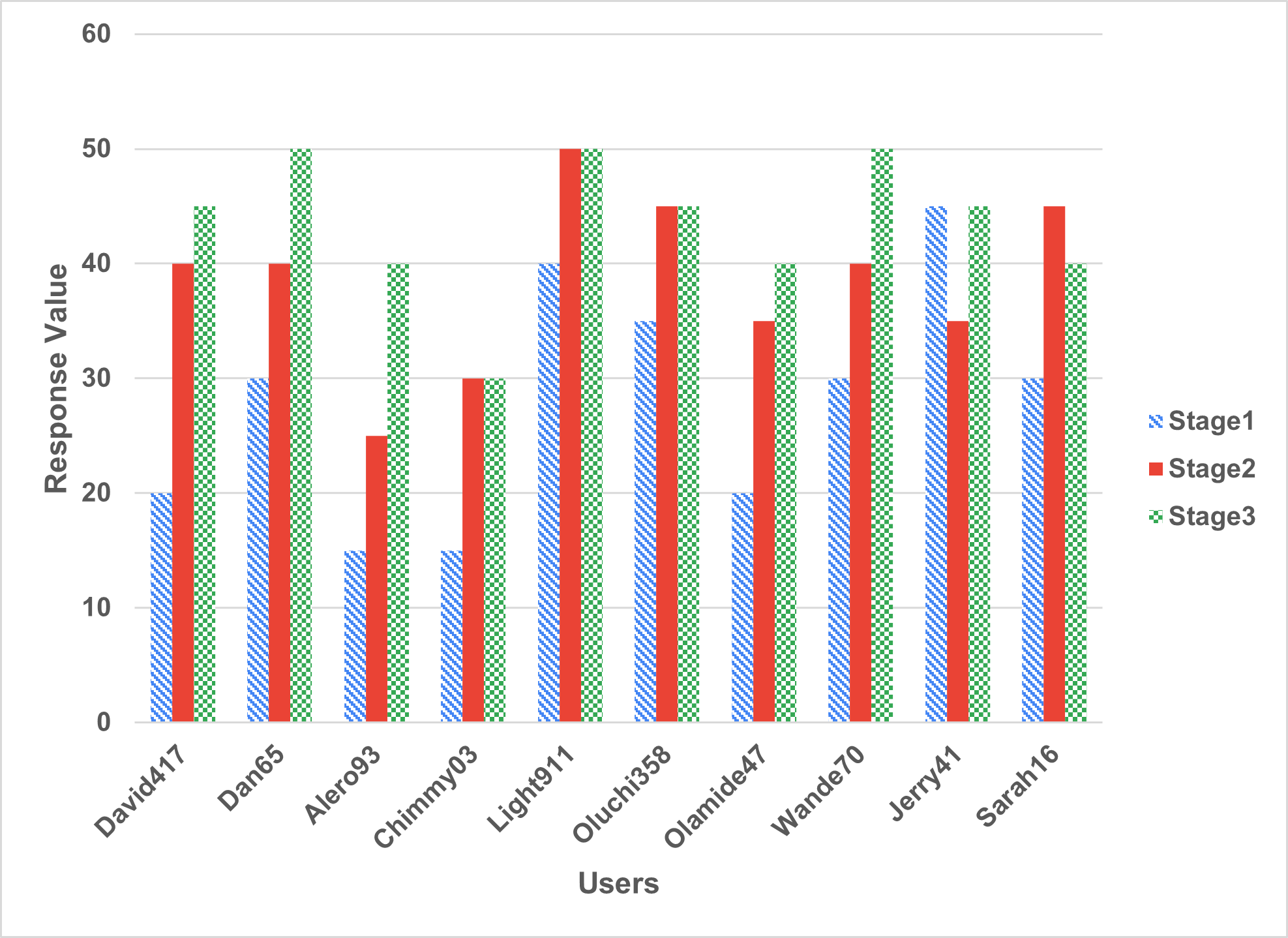}}}\hfill

\subfloat[No. of Hints taken]{\label{fig:6b}{\includegraphics[width=.6\textwidth]{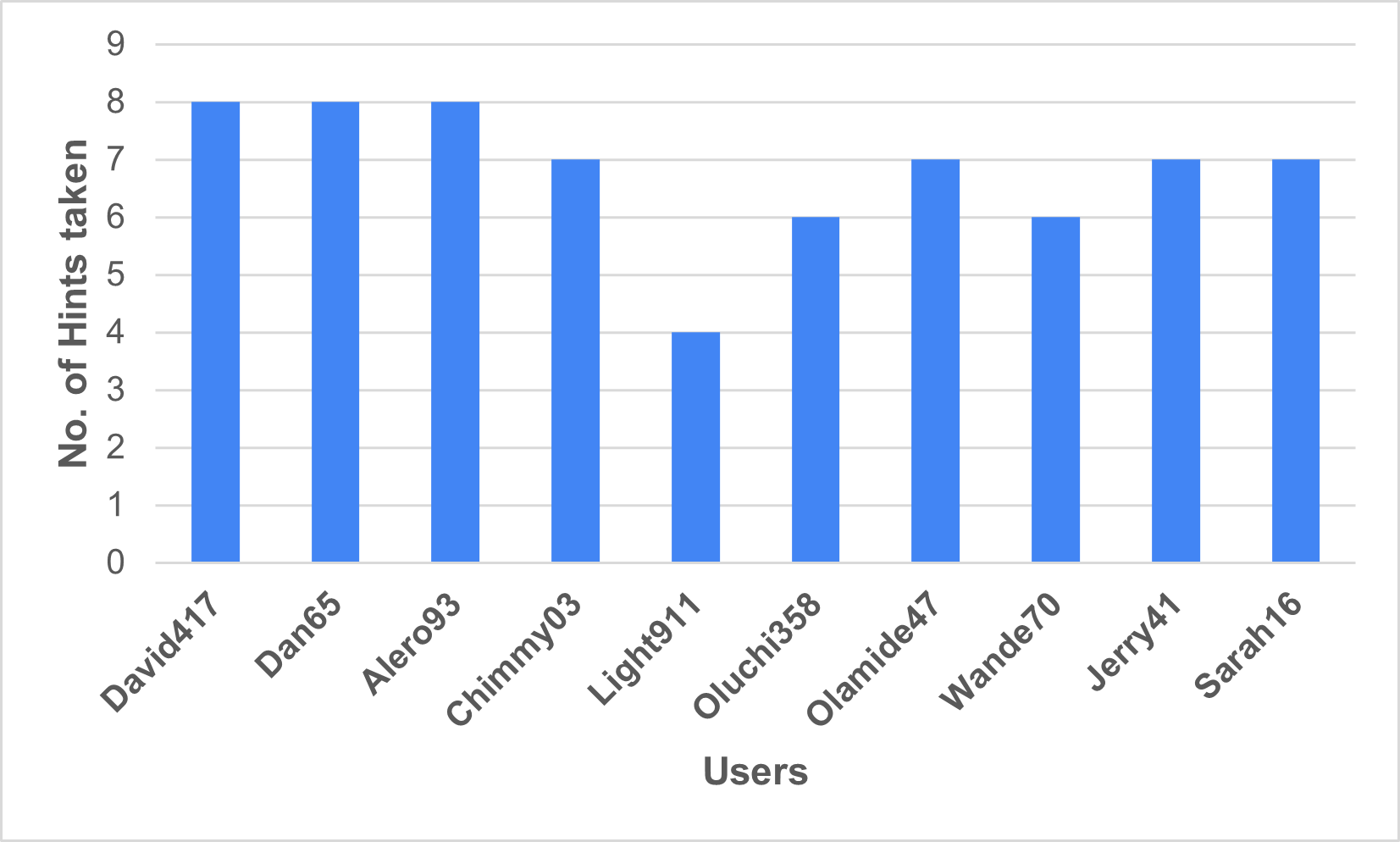}}}\hfill


\caption{Intermediate level Evaluation of the designed system.}
\label{fig:6}
\end{figure*}

As can be seen from Fig. \ref{fig:6a}, all ten players in the intermediate level performed well at all stages. The maximum score that can be earned is 50; however, the scores obtained by the players vary, which may be attributed to the level of understanding regarding smart meter security of the players prior to participating in the training. Observe that Jerry41 obtained a 45/50 score, suggesting that the player already has a good understanding of smart meter security and vulnerability, and it would be difficult to observe the impact of training. Alero93 and Chimmy03 obtained the lowest scores with 15 marks each, followed by Olamide47 and David17 with 20 marks each.

Fig. \ref{fig:6b} Illustrates that the number of hints available to the players influenced their performance in learning the answers to the questions. There is no doubt that most of the players performed better with hints than without them, and since the hint shows high engagement, then it can be concluded that most of the players relied on it to perform better. Light911 took the last hint, thus showing that he has a good understanding of the smart meter system vulnerability. He scored 40 points in Stage One and 50 points in Stage Two. Among all participants in this stage, Alero93 received the lowest mark with 25 marks, which was also the result of his performance in stage One. Despite taking all of the hints, Alero93 only made a slight improvement, and the same was true for chimmy03, who had the same score in stage One as Alero93.
As can be clearly seen, all players performed better in stage three except for Sarah16, who scored 30 in stage one, 45 in stage two, and 40 in stage three. David417's performance at stage two shows the most significant impact of the training. David417 scored 20 in stage one, took all the hints, and then improved to 40 in stage two and 45 in stage three. After scoring 30 in Stage One, Wande70 took six hints to reach 40 in Stage Two, and his performance then increased to 50 in Stage Three utilizing hints from Stage Two.

\subsubsection{Advanced-Level Evaluation}

\begin{figure*}[]
\centering

\subfloat[User's score of the three stages]{\label{fig:7a}{\includegraphics[width=.6\textwidth]{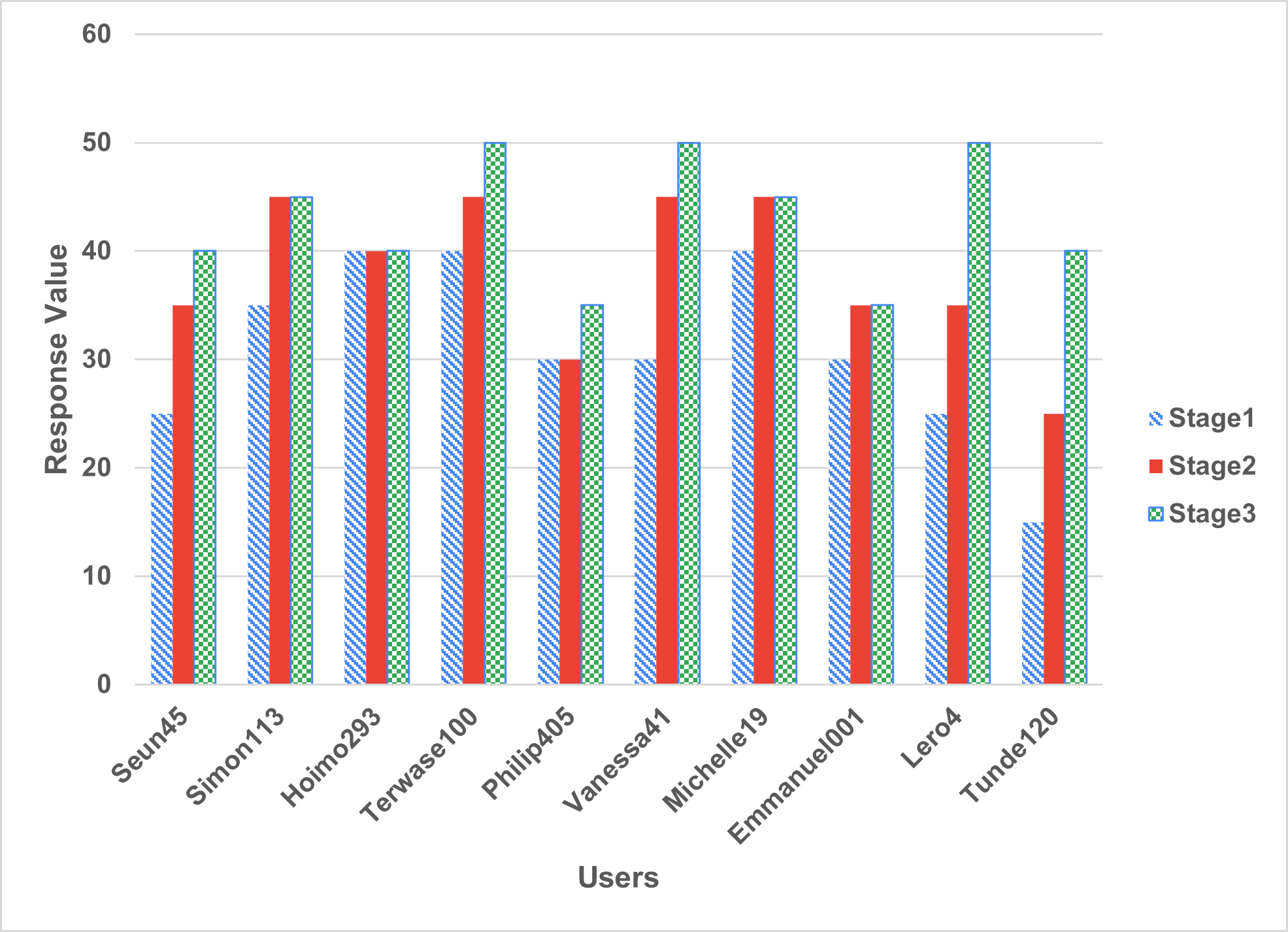}}}\hfill

\subfloat[No. of Hints taken]{\label{fig:7b}{\includegraphics[width=.6\textwidth]{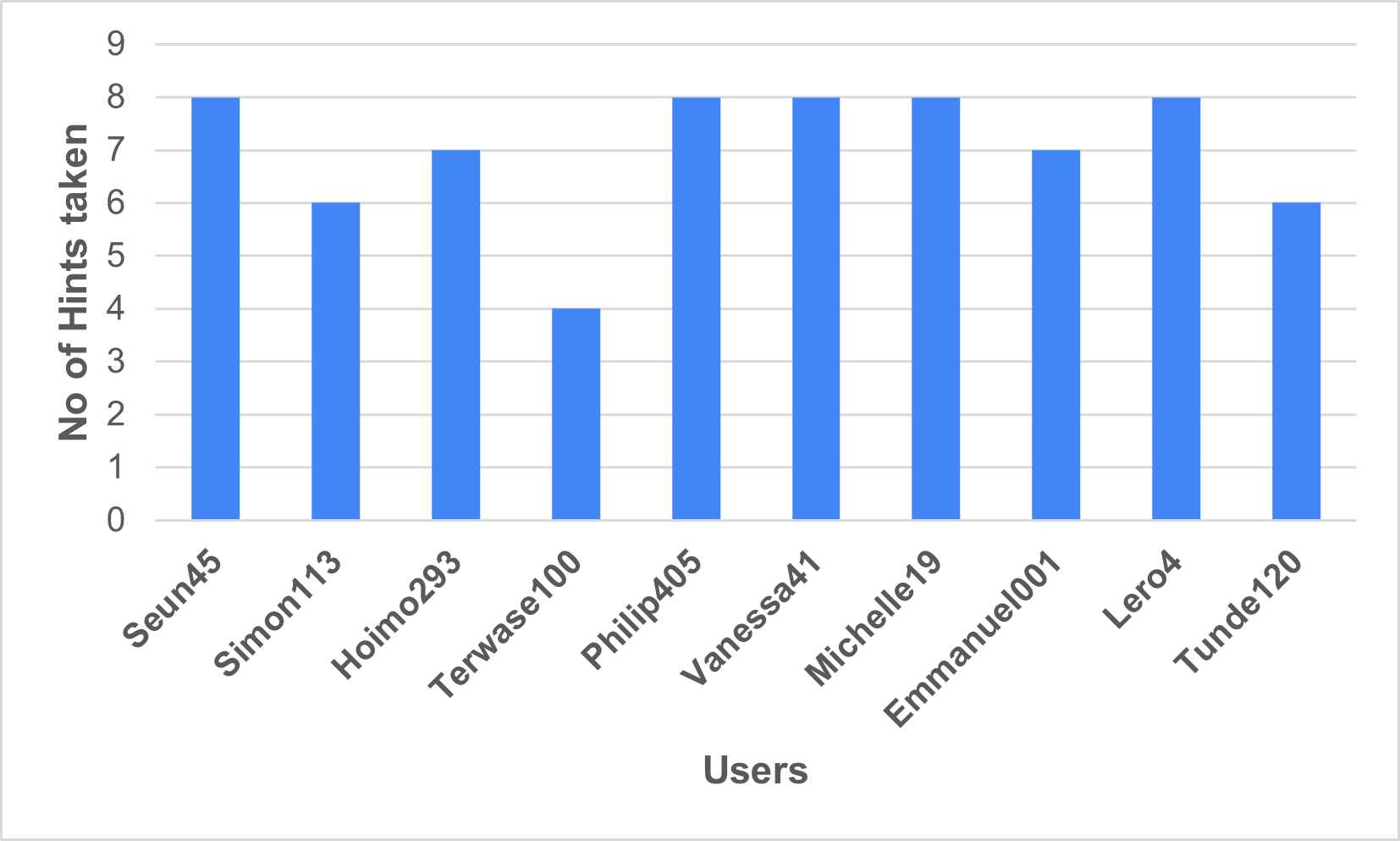}}}\hfill

\caption{Advanced level Evaluation of the designed system. }
\label{fig:7}
\end{figure*}

Fig. \ref{fig:7a} shows the score distribution of the advanced players in the three stages. It appears that some players performed very well, which indicates that they had a good understanding of the subject prior to the training. There are, however, some who performed moderately and others who performed relatively poorly. This stage has the highest score of 40/50, achieved by three individuals (Hoimo, Terwase, and Michelle). Tunde received the lowest score of 15/50. 
During the second stage, hints are provided to the students so that they can evaluate the improvement in their performance, as shown in Fig \ref{fig:7b}. Observation reveals that all players took the hints to improve their knowledge. Thus the performance can be interpreted as reflecting the impact of the training received. Considering Terwase's high performance in stage One and stage Two, it can be argued that she had an understanding of the system prior to the training. As a result, Tunde's score in Stage 2 increased from 15 to 25 after he received the lowest score in Stage One. Although she took eight hints, the small forward also improved from 25 to 35. 

Based on the training hints provided in stage Two, some players were able to improve their performance. Terwase, Vanessa, and Alero each had a 50/50. While Terwase's score cannot be assumed to have been considerably impacted by the hints because he has maintained a high score since the phase, Vanessa and Alero improved dramatically from 30/50 in Stage One to 50/50 in Stage Three and from 25/50 to 50/50, respectively. It can be observed that Philip and Emmanuel recorded the lowest scores in stage three, as their scores indicate that the hints taken in the second stage had little impact on their scores. Both of them took eight and seven hints, respectively, in the second stage. Considering this, it may be assumed that the two participants were not aware of the training since the scores of all other participants were improved as a result of the hints.

\subsection{Overall Flag Captured}

\begin{figure}[]
\centering

{\includegraphics[width=.48\textwidth]{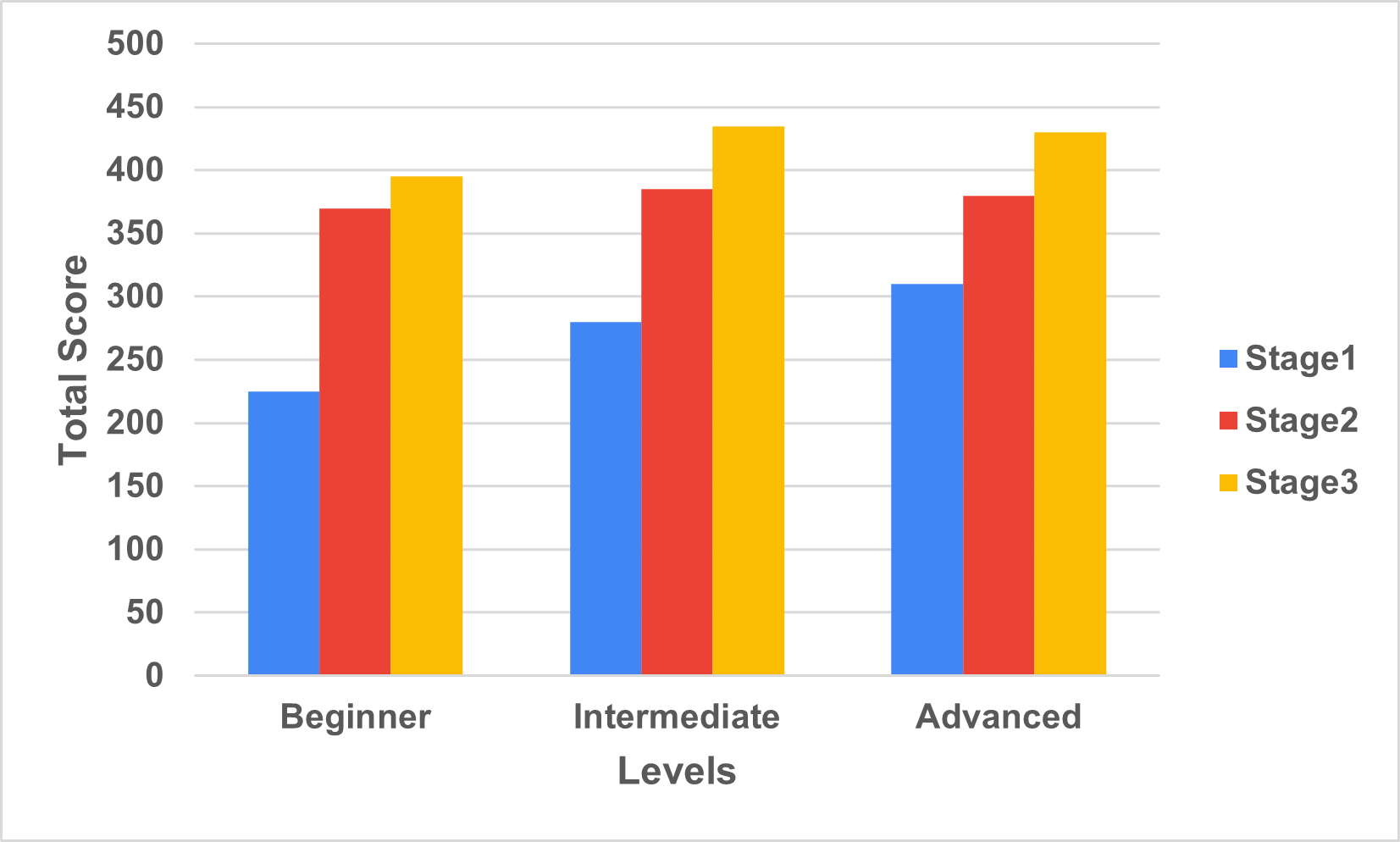}}

\caption{Overall average progression in performance across three levels.}
\label{fig:9}
\end{figure}

Table \ref{tab2} and Fig. \ref{fig:9} show the overall flag captured by players on each of the questions attempted. It can be observed that in the third stage of the beginner level, the question area with the least capture rate is device security, HTTP and data breach (having 6 successful captures). Based on these findings, we can conclude that participants will have a relatively low level of knowledge in this area and will require more training to become proficient. They are, however, already proficient in the implementation of SMS links, physical security, and authentication (with a total of 10, 9 and 9 successful captures, respectively). Users were prompted with questions regarding a variety of security topics related to smart meters, including physical security, passwords, phishing, cyber-calling, and public Wi-Fi.

The performance at the intermediate level stage 3 was quite good, with the least successful capture being 7 on questions related to system updates. Other areas have good performance, indicating that participants have been well-trained using the approach and now have a better understanding of the smart grid vulnerabilities at an intermediate level, thanks to gamification. This level of questioning covered a variety of security topics related to smart meters, including network security, USB safety, privacy, and e-mail scams.
The performance has also been significantly improved in stage 3 of the advanced level, where the least successful capture rate on questions related to data injection is 7. Gamification has also been used extensively in other areas. There were various security topics related to smart meters at this level, such as smart infrastructure, microcontrollers, denial of service attacks, eavesdropping, data injection, etc. 
The first stage evaluates the payers' initial level of awareness prior to the training, the second stage assesses their performance during training, and the third stage measures their performance following training. As evidenced by Fig. \ref{fig:9}, the study's approach led to an apparent increase in the level of awareness among the participants.



\begin{table*}[]
\centering
\caption{Overall Capture of all questions within the three levels.}
\label{tab2}
\begin{tabular}{|l|c|c|c|}
\hline
\textbf{User name} & \textbf{Capture rate (\%)} & \textbf{Successful Captures} & \textbf{Failed attempt} \\
\hline
\multicolumn{4}{|c|}{\textbf{Beginner Level}} \\
\hline
Physical Security	& 28	& 9	& 1 \\
\hline
Passwords	& 21	& 7	& 3\\
\hline
Phishing	& 25	& 8	& 3\\
\hline
Cyber-Calling	& 21	& 7	& 3\\
\hline
Public Wi-FI	& 25	& 8	& 2\\
\hline
Authentication	& 28	& 9	& 1\\
\hline
Device Security	& 18	& 6	& 5\\
\hline
Http	& 18	& 6	& 4\\
\hline
Data Breach	& 18	& 6	& 4\\
\hline
SMS Links	& 31	& 10	& 1\\
\hline
\multicolumn{4}{|c|}{\textbf{Intermediate Level}} \\
\hline
Network Security &	31	& 10	& 0 \\
\hline
Password Security	&	25	& 8	& 2\\
\hline
Malware	&	25	& 8	& 2\\
\hline
Authentication	&	25	& 8	& 2\\
\hline
System Updates	&	21	& 7	& 3\\
\hline
Email Scam	&	28	& 9	& 1\\
\hline
USB Safety		& 25	& 8	& 2\\
\hline
Publishing Updates	&	31	& 10	& 0\\
\hline
Data	&	28	& 9	& 1\\
\hline
Privacy	&	31	& 10	& 1\\
\hline
\multicolumn{4}{|c|}{\textbf{Advanced Level}} \\
\hline
Network Security	&	31	& 10	& 0 \\
\hline
Smart Infra	&	25	& 8	& 2\\
\hline
Hardware	&	28 &	9	& 2\\
\hline
Microcontroller	& 	31 &	10	& 0\\
\hline
Data	&	28 &	9 &	2\\
\hline
DoS	&  	28	& 9	 & 1\\
\hline
Access Request	&	25	& 8	& 3\\
\hline
Eavesdropping	&	25	& 8	& 2\\
\hline
Data Injection	&	21	& 7	& 4\\
\hline
Remote Login	&	25	& 8	& 2\\
\hline

\end{tabular}
\end{table*}


\section{Future research directions\label{sec5}}

The study demonstrated a critical analysis of the impact of gamification on security awareness training, the system prototype which was developed to evaluate the performance of the players had a few challenges that require necessary review that must be considered should any progressive study in this area be proposed in the future. This challenge led to the limited number of participants, perhaps the result could have been more accurate if the sample size is increased. In view of these, the following insight was recommended for future engagement of similar nature:

\begin{enumerate}
    \item The gaming system should be hosted on a server that allows for the sharing of links and individual access to the game, thereby eliminating geographical limitations on the pool of potential respondents and participants.

    \item To enhance the efficiency of security awareness education, a personality test questionnaire should be included in the evaluation to gain insight into the participants' perceptions and traits.

    \item Increasing the number of questions that can be selected for further examination of security concepts is possible. Furthermore, certain scenarios can be targeted within those questions.

    \item Creating immersive and interactive gaming experiences for security training could improve awareness by utilizing virtual reality technology.

    \item An evaluation of the long-term effects of gamified security awareness training on user behaviour and behaviour. Although some studies have been conducted on the impact of gamified SAT, there is still a need to emphasize the behaviours and actions of the users.
    
    \item Ensure that users receive regular training and feedback on the latest security threats and best practices by providing them with up-to-date content.

\end{enumerate}

\section{Conclusion\label{sec6}}

Gamified learning is examined in this paper with regard to knowledge retention. In order to assess participants' knowledge and improvement with gamification, a prototype system was developed using root the box. As a result of gamification, the results show an immediate improvement in performance as well as a greater level of participation when hints are provided. It can be demonstrated that the scores of Participants in the three levels have improved by 40\%, 35\% and 29\%, respectively. This result reflects the awareness of learning within our system. Also, we have calculated the number of hints used in the system that represents the participants' learning ability. Moreover, this was evident in the study as participants, especially those who scored low in stage 1, were encouraged that they could be better only if they had a hint to remind them. Consequently, the study has demonstrated the effectiveness of the approach and its applicability to a number of other application areas, including cybersecurity.

\vspace{6pt}
\textbf{Authors Contribution:} Conceptualization, Y.A.; methodology, Y.A.and M.E; validation, Y.A. and M.E.; formal analysis, Y.A. and M.E; investigation, Y.A.; and M.E; resources, Y.A. and M.E.; M.A; H.M.; and M.Y.; data curation, Y.A. and M.E; writing---original draft preparation, Y.A.; writing---review and editing, Y.A.; M.E.; and H.M; visualization, Y.A., M.E; and H.M;  supervision, Y.A.; project administration, Y.A.; M.A.; H.M.; and M.Y.; funding acquisition, Y.A; M.A. All authors have read and agreed to the published version of the manuscript.

\vspace{6pt}

\textbf{Funding:} No funding was received to assist with the preparation of this paper.

\vspace{6pt}

\textbf{Data availability:}: Dataset available on request from the authors.
\vspace{6pt}

\textbf{Conflict of interest:} The authors have no conflicts of interest to declare that are relevant to the content of this paper.

\vspace{6pt}




\bibliographystyle{unsrt}

\bibliography{main}

\end{document}